\documentclass[twocolumn,superscriptaddress,showpacs,amssymb,preprintnumbers,prl,floatfix]{revtex4-1}
\usepackage{epsfig,dcolumn}
\usepackage{hyperref}
\usepackage{multirow}
\usepackage{bm,color}
\usepackage{tikz}
\usepackage{makecell}
\usepackage{graphicx,bm}

\newcommand{\be}{\begin{equation}}
\newcommand{\ee}{\end{equation}}
\newcommand{\ba}{\begin{eqnarray*}}
\newcommand{\ea}{\end{eqnarray*}}

  \usepackage{ulem}

\usepackage{soul}
\newcommand{\Capd}{$^{37}$Ca($p$,$d$)$^{36}$Ca}
\newcommand{\Capt}{$^{38}$Ca($p$,$t$)$^{36}$Ca}

\begin{document}
\title{The structure of $^{36}$Ca under the Coulomb magnifying glass}

\author{L.~Lalanne}
    \email[]{present adress: louis-alexandre.lalanne@cern.ch}
    \affiliation{Universit\'e Paris-Saclay, CNRS/IN2P3, IJCLab, 91405 Orsay, France}
    \affiliation{Grand Accélérateur National d'Ions Lourds (GANIL), CEA/DRF-CNRS/IN2P3, 
    Bd. Henri Becquerel, 14076 Caen, France}  
\author{O.~Sorlin}
    \email[]{olivier.sorlin@ganil.fr}
    \affiliation{Grand Accélérateur National d'Ions Lourds (GANIL), CEA/DRF-CNRS/IN2P3, 
    Bd. Henri Becquerel, 14076 Caen, France}
    \author{A.~Poves}
    \affiliation{Departamento de F\'isica Te\'orica and IFT-UAM/CSIC, Universidad Aut\'onoma de Madrid,  E-2804 Madrid, Spain}
\author{M.~Assi\'e}
    \affiliation{Universit\'e Paris-Saclay, CNRS/IN2P3, IJCLab, 91405 Orsay, France}
\author{F.~Hammache}
    \affiliation{Universit\'e Paris-Saclay, CNRS/IN2P3, IJCLab, 91405 Orsay, France}
\author{S.~Koyama}
    \affiliation{Department of Physics, Unviversity of Tokyo}
     \affiliation{Grand Accélérateur National d'Ions Lourds (GANIL), CEA/DRF-CNRS/IN2P3, 
    Bd. Henri Becquerel, 14076 Caen, France}
\author{D.~Suzuki}
    \affiliation{RIKEN Nishina Center, 2-1, Hirosawa, Wako, Saitama 351-0198, Japan}
\author{F.~Flavigny}
    \affiliation{Normandie Univ, ENSICAEN, UNICAEN, CNRS/IN2P3, LPC Caen, 14000 Caen, France}
\author{V.~Girard-Alcindor}
    \affiliation{Grand Accélérateur National d'Ions Lourds (GANIL), CEA/DRF-CNRS/IN2P3, 
    Bd. Henri Becquerel, 14076 Caen, France}
    \author{A.~Lemasson}
    \affiliation{Grand Accélérateur National d'Ions Lourds (GANIL), CEA/DRF-CNRS/IN2P3, 
    Bd. Henri Becquerel, 14076 Caen, France}
  \author{A.~Matta}
       \affiliation{Normandie Univ, ENSICAEN, UNICAEN, CNRS/IN2P3, LPC Caen, 14000 Caen, France}
       \author{T.~Roger}
    \affiliation{Grand Accélérateur National d'Ions Lourds (GANIL), CEA/DRF-CNRS/IN2P3, 
    Bd. Henri Becquerel, 14076 Caen, France}
\author{D.~Beaumel}
    \affiliation{Universit\'e Paris-Saclay, CNRS/IN2P3, IJCLab, 91405 Orsay, France}
\author{Y~Blumenfeld}
    \affiliation{Universit\'e Paris-Saclay, CNRS/IN2P3, IJCLab, 91405 Orsay, France}
\author{B.~A.~Brown}
 \affiliation{Department of Physics and Astronomy, National Superconducting Cyclotron Laboratory,
Michigan State University, East Lansing, Michigan}
\author{F.~De Oliveira Santos}
 \affiliation{Grand Accélérateur National d'Ions Lourds (GANIL), CEA/DRF-CNRS/IN2P3,
    Bd. Henri Becquerel, 14076 Caen, France}
\author{F.~Delaunay}
    \affiliation{Normandie Univ, ENSICAEN, UNICAEN, CNRS/IN2P3, LPC Caen, 14000 Caen, France}
    \author{N.~de~S\'{e}r\'{e}ville}
    \affiliation{Universit\'e Paris-Saclay, CNRS/IN2P3, IJCLab, 91405 Orsay, France}
\author{S.~Franchoo}
    \affiliation{Universit\'e Paris-Saclay, CNRS/IN2P3, IJCLab, 91405 Orsay, France}
\author{J.~Gibelin}
    \affiliation{Normandie Univ, ENSICAEN, UNICAEN, CNRS/IN2P3, LPC Caen, 14000 Caen, France}
\author{J.~Guillot}
    \affiliation{Universit\'e Paris-Saclay, CNRS/IN2P3, IJCLab, 91405 Orsay, France}
\author{O.~Kamalou}
    \affiliation{Grand Accélérateur National d'Ions Lourds (GANIL), CEA/DRF-CNRS/IN2P3, 
    Bd. Henri Becquerel, 14076 Caen, France}
\author{N.~Kitamura}
    \affiliation{Center for Nuclear Study, University of Tokyo}

\author{V.~Lapoux}
    \affiliation{CEA, Centre de Saclay, IRFU, Service de Physique Nucléaire, 91191 Gif-sur-Yvette, France}

\author{B.~Mauss}
    \affiliation{RIKEN Nishina Center, 2-1, Hirosawa, Wako, Saitama 351-0198, Japan}
    \affiliation{Grand Accélérateur National d'Ions Lourds (GANIL), CEA/DRF-CNRS/IN2P3, 
    Bd. Henri Becquerel, 14076 Caen, France}
\author{P.~Morfouace}
    \affiliation{Grand Accélérateur National d'Ions Lourds (GANIL), CEA/DRF-CNRS/IN2P3, 
    Bd. Henri Becquerel, 14076 Caen, France}
    \affiliation{CEA, DAM, DIF, F-91297 Arpajon, France}
\author{M.~Niikura}
    \affiliation{Department of Physics, Unviversity of Tokyo}
\author{J.~Pancin}
    \affiliation{Grand Accélérateur National d'Ions Lourds (GANIL), CEA/DRF-CNRS/IN2P3, 
    Bd. Henri Becquerel, 14076 Caen, France}

\author{T.~Y.~Saito}
    \affiliation{Department of Physics, Unviversity of Tokyo}

\author{C.~Stodel}
    \affiliation{Grand Accélérateur National d'Ions Lourds (GANIL), CEA/DRF-CNRS/IN2P3, 
    Bd. Henri Becquerel, 14076 Caen, France}
\author{J-C.~Thomas}
    \affiliation{Grand Accélérateur National d'Ions Lourds (GANIL), CEA/DRF-CNRS/IN2P3, 
    Bd. Henri Becquerel, 14076 Caen, France}

\begin{abstract}

Detailed spectroscopy  of the neutron-deficient nucleus $^{36}$Ca was obtained up to 9 MeV using the $^{37}$Ca($p$,$d$)$^{36}$Ca and the $^{38}$Ca($p$,$t$)$^{36}$Ca transfer reactions. 
The radioactive nuclei, produced by the LISE spectrometer at GANIL, interacted with the protons of
the liquid Hydrogen target CRYPTA, to produce light ejectiles (the deuteron $d$ or
triton $t$) that were detected in the MUST2 detector array, in coincidence with the heavy residues  identified by a zero-degree detection-system. Our main findings are: i) a similar shift in energy for the 1$^+_1$ and 2$^+_1$ states by about -250 keV, as compared to the mirror nucleus $^{36}$S,  ii) the discovery of an intruder  0$^+_2$  state at 2.83(13) MeV, which appears below the first 2$^+$ state, in contradiction with the situation in $^{36}$S, and iii) a tentative 0$^+_3$ state at 4.83(17) MeV, proposed to exhibit a bubble structure with two neutron vacancies in the 2s$_{1/2}$ orbit. The inversion between the 0$^+_2$ and 2$^+_1$ states is due to the large mirror energy difference (MED) of -516(130) keV for the former. 
This feature is reproduced by Shell Model (SM) calculations, using the $sd$-$pf$ valence space, predicting an almost pure intruder nature for the 0$^+_2$ state, with two protons (neutrons) being excited across the $Z$=20 magic closure in $^{36}$Ca ($^{36}$S).  This mirror system has the largest MEDs ever observed, if one excludes the few cases induced by the effect of the continuum.

 \end{abstract}

\keywords{Proton rich nuclei, Coulomb effects, Shell Model,
 $sdpf$-shell spectroscopy, Level schemes and transition probabilities.}

\date{\today}
\maketitle

\noindent {\sl Introduction.} 
The studies of fundamental symmetries and the mechanisms that induce their breaking are crucial to the understanding and appreciation of the wealth of the physics processes ruling our world \cite{Lee56,Higgs,Wein96}.  In atomic nuclei, the isospin symmetry is born out of the charge independence of the strong interaction which considers that protons and neutrons are two representations of the same particle, the nucleon. The electromagnetic interaction violates this symmetry and is the main mechanism responsible for isospin symmetry breaking effects (ISB). However, even if the Coulomb contribution to the total binding energy of the nucleus is quite large, it barely affects its spectroscopic properties and energy-level schemes of mirror nuclei (with interchanged numbers of protons and neutrons) are generally found to be nearly identical. 

ISB are known to produce small differences in the excitation energies of analogue states in a pair of mirror nuclei, which are dubbed Mirror Energy Differences (MEDs) \cite{Nolen69,TE1,TE2}. The difference in $E2$ transition matrix elements between mirror nuclei has also been used as a probe of ISB, see e.g., \cite{Wimm21,Boso19,Gil19}. The Coulomb repulsion among the protons is the main source of MED. Its amplitude  is generally small (10-100 keV) and very rarely exceeds $\pm$~200~keV \cite{henderson20}. However, even a small MED of only few tens of keV can produce quite a  prominent effect, such as different ground-state spin values between the mirror pair $^{73}$Sr-$^{73}$Br~\cite{Hoff20}, commented in ~\cite{henderson20, Lenzi20}. The study of MED probes in a unique manner the wave function of the nucleons inside the nucleus. Remarkable examples are (i) the evolution of MED along rotational bands, which yields insight into the changes in spatial correlations and spin alignment \cite{War06,Lenzi01,Bentley06}, (ii) the large MED of up to 700 keV observed  in the  $A$ = 13 ($^{13}$C - $^{13}$N) \cite{TE1,TE2} and $A$ = 16 ($^{16}$N - $^{16}$F) mirror pairs \cite{Stef14}, also called Thomas-Ehrmann shifts, which probes the spatial expansion of unbound $s$-orbits and the influence of the continuum, (iii) the persistence of mirror symmetry in the disappearance of the magic number 8 between $^{12}$O and its partner $^{12}$Be \cite{Suzu09, Suzu16}
and (iv) a proposed change of shape between the mirror nuclei  $^{70}$Kr and $^{70}$Se\cite{Wimm21}, based on $E2$ reduced transition matrix elements, which is however questioned in \cite{Lenz21}.

In the present work, we provide for the first time evidence of very large MED in conjunction with the phenomenon of shape coexistence through the experimental and theoretical studies of the 0$^+_2$, 2$^+_1$ and 1$^+_1$ states in the 
$A=36$, $T=2$ mirror pair, $^{36}$S and  $^{36}$Ca. 
 Note that these states are not subject to TE shifts as the Coulomb barrier of $^{36}$Ca ($\simeq 6.1$ MeV) is much higher than the one- and two-proton emission thresholds ($\simeq 2.6$ MeV and
2.68 MeV, respectively). 

\noindent {\sl Experimental techniques.}  The $^{37}$Ca and $^{38}$Ca nuclei were produced at about 50 MeV/nucleon by fragmentation reactions of a 95~MeV/nucleon $^{40}$Ca$^{20+}$ beam, with an average intensity of $\sim$2~$\mu$Ae, in a 2-mm thick $^{9}$Be target. They were selected through two different settings of the LISE3/GANIL spectrometer~\cite{Ann}, leading to a purity of 20\% and mean rates of 3$\times$10$^3$ pps and 2$\times$10$^4$ pps, respectively.  They were subsequently tracked 
by two low-pressure multi-wire detectors, CATS \cite{cats}, before interacting with protons of a cryogenic liquid Hydrogen target CRYPTA \cite{Koy20} (of effective thickness of 9.7 mg$\,$cm$^{-2}$).  They were unambiguously identified by means of their time-of-flight (TOF) measurement between the CATS detectors and the cyclotron radio frequency.

The outgoing ions were detected by a Zero Degree Detection (ZDD) system, composed of an ionization chamber (IC), yielding their $Z$ identification, a set of two $XY$ drift chambers (DC), used to determine their outgoing angles, and a thick plastic scintillator, mostly used for time-of-flight measurements. The energy and angle of the light outgoing particles, either $d$ or $t$ from the transfer reactions, as well as proton(s) emitted from unbound states, were measured by a set of six MUST2 telescopes \cite{must2}, each composed of a first stage of a 300-$\mu$m thick double-sided Silicon strip detector (DSSSD) and a second stage of sixteen 4-cm thick CsI crystals. Light particle identification was performed using the correlation between the energy loss, $\Delta E$, and the residual energy, $E$, measured in the DSSSD and the CsI crystals, respectively (see Refs.~\cite{Lal21, PhD} for more details).

\begin{figure}[t]
\centering

\includegraphics[width=0.99\columnwidth]{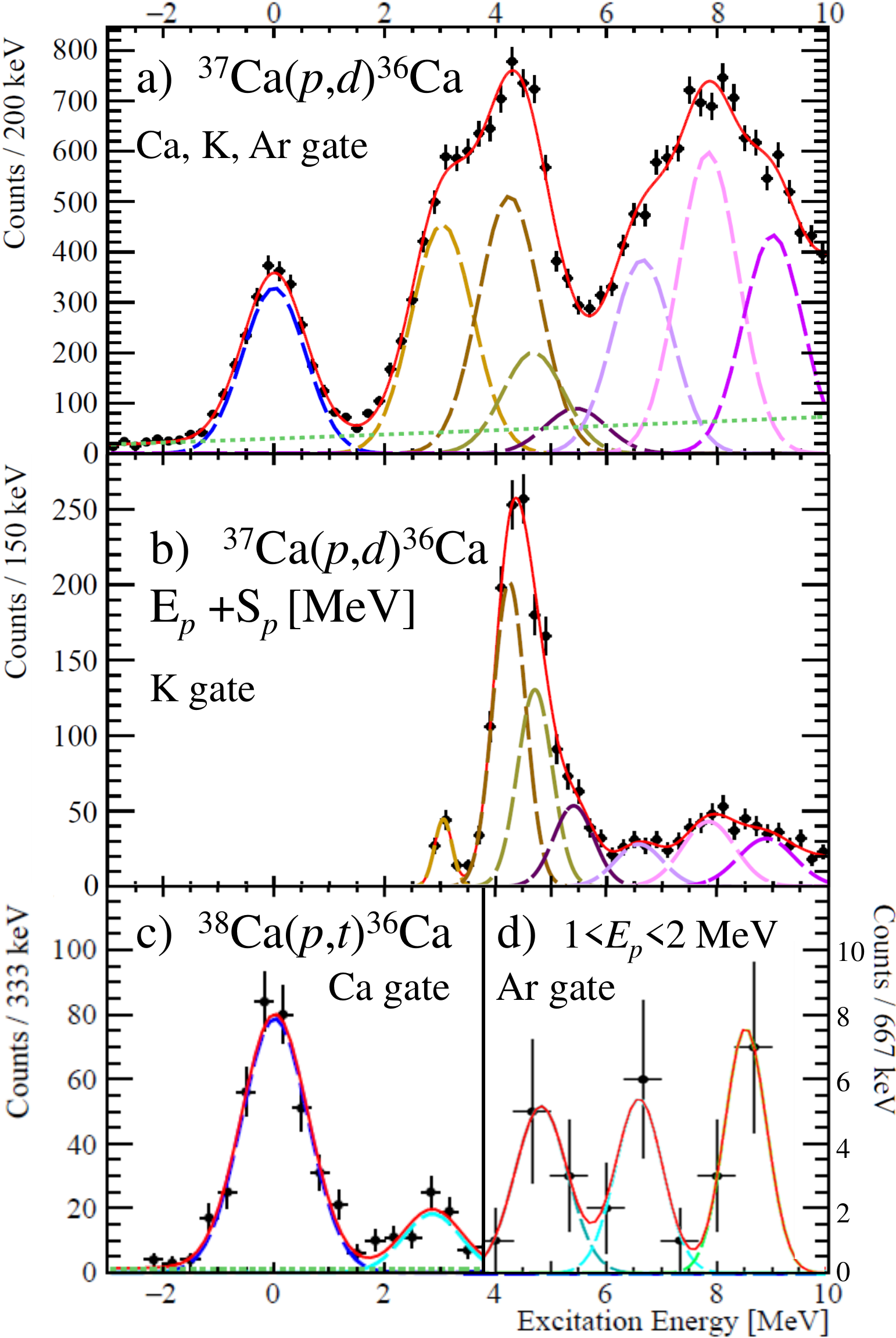}
\caption{ $a)$ Excitation energy spectrum $E_x$ of $^{36}$Ca obtained from the $(p,d)$ transfer reaction with a gate on outgoing Ca, K and Ar.  $b)$ One-proton energy spectrum  (to which S$_p$($^{36}$Ca)$=2599.6(61)$~keV has been added) obtained from the $(p,d)$ reaction with a gate on outgoing K isotopes. $E_x$ spectrum from the $(p,t)$ reaction and gated on outgoing Ca is shown in $c)$, while the one in $d)$ is gated on outgoing Ar and the detection of one proton with a center-of-mass energy between 1 and 2 MeV.  Individual contributions to the fits are shown with different color codes. }
\label{fig:Ex}
\end{figure} 

\noindent {\sl Results.} The excitation energy $E_x$ of $^{36}$Ca was obtained from the $^{37}$Ca($p$,$d$)$^{36}$Ca (\Capt) reactions using the missing mass method after gating on an incoming $^{37}$Ca ($^{38}$Ca) in CATS and on a $d$ ($t$) particle in MUST2. The total excitation energy spectrum obtained using the ($p$,$d$) reaction and gated on outgoing Ca, K and Ar isotopes (using the ZDD) is shown in Fig.~\ref{fig:Ex}a). The spectra obtained from the ($p$,$t$) reaction,  gated  only on outgoing Ca or only Ar, are shown in Figs.~\ref{fig:Ex}c) and d), respectively.  Fig~\ref{fig:Ex}b) shows the one-proton energy reconstructed in the center-of-mass of $^{36}$Ca using the ($p$,$d$) reaction with a gate on outgoing K only. Since the energy resolution reconstructed with the protons is better than with the deuterons (\cite{Lal21, PhD}), Fig~\ref{fig:Ex}b) allows the determination of the energy peak centroids to be used for the fit of Fig~\ref{fig:Ex}a) in the 4-5.5~MeV excitation energy range.

The red lines in Figs. \ref{fig:Ex} a-d) display the best fits obtained using multiple Gaussian functions (shown with colored dotted lines) plus a small background contribution (green dashed line), generated by interactions of the beam particles with the windows of the LH$_2$ target, determined in a dedicated run with an empty target. The width of each peak used in the fit is constrained by simulations performed with the $nptool$ package~\cite{Matta2016}, the reliability of which is checked from the observed widths of isolated peaks (e.g., 0$^+_1$ at 0~MeV in Fig.~\ref{fig:Ex}a) and 2$^+_1$ at 3~MeV in Fig.~\ref{fig:Ex}b)). A typical energy resolution in $E_x$ of about 550~keV  is found for the peaks of Figs.~\ref{fig:Ex}a,c,d). The resolution in proton energy varies from 130~keV at  3~MeV to 500~keV at 8~MeV in Fig.~\ref{fig:Ex}b).  The number of contributions used in the fit was guided by the statistical tests of the $\chi^2$ and the $p$-value, as well as the number of states populated in the quasi mirror reaction $^{37}$Cl($d$,$^3$He)$^{36}$S \cite{Gray70}. Finally, contributions of the one and two-proton phase spaces have been found to be of less than 2\% for excitation energies below 10~MeV ~\cite{Suppl1}. 

\begin{figure}[!h]
\begin{center}
\centering

\includegraphics[width=0.99\columnwidth]{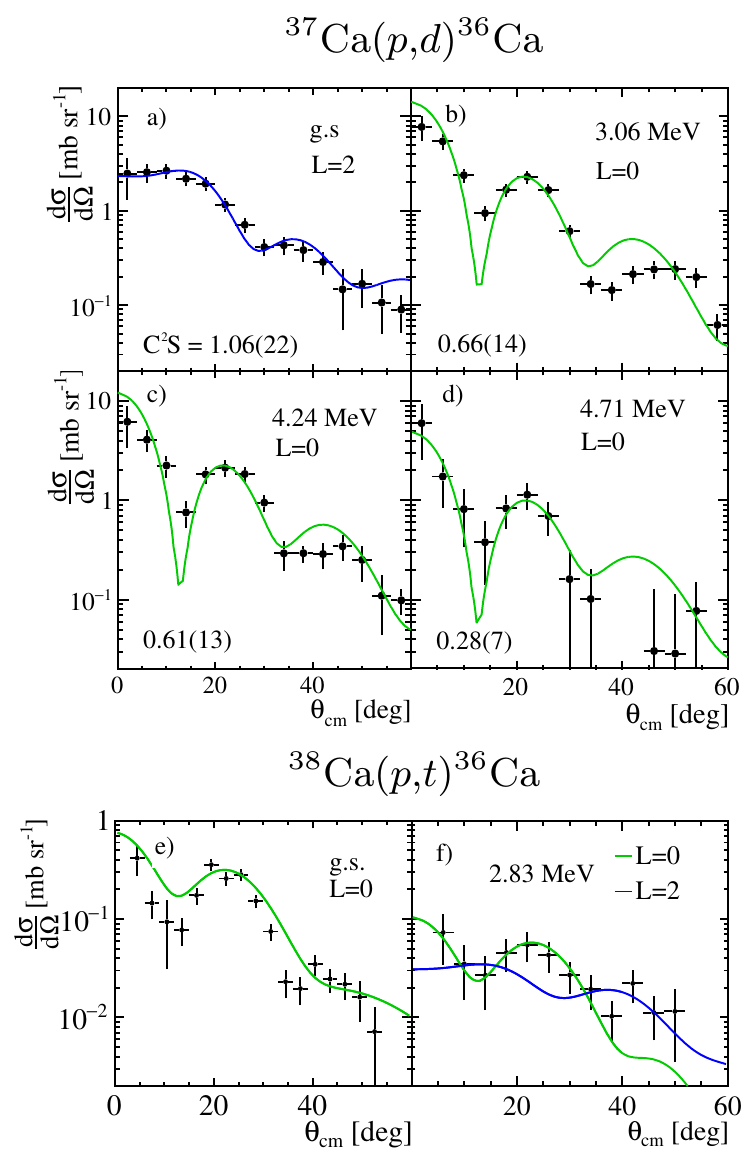}

 \end{center}
\caption{Differential cross sections of the states identified in $^{36}$Ca using  the \Capd~ and  \Capt~ transfer reactions are shown in a-d) and e,f), respectively. Blue (green) full lines of panels a-d) show the results of the $L=2$ ($L=0$) DWBA calculations fitted to the data. In panels e,f), the green and blue full lines correspond to results of the DWBA plus TNA calculations  assuming $L=0$ and $L=2$ transfers, respectively. Panel e) used  TNA  values directly extracted from the present shell-model calculations, while panel f) is a result of fitted values (see text for details).}
\label{fig:XS}
\end{figure}

The  differential cross sections corresponding to the \Capd~ and \Capt~ transfer reactions are shown in Figs.~\ref{fig:XS}a-d) and ~\ref{fig:XS}e,f), respectively. They have been obtained by fitting the excitation energy spectra of Fig.~\ref{fig:Ex}a,c) (with centroid values fixed at the energies measured using the full angular range) by slices of center-of-mass angles. A proper normalization of their amplitudes is made using the number of incident nuclei, the density of target protons, as well as taking into account the geometrical and detection efficiencies of the experimental setup. The shape of the distributions gives clear information on  the transferred angular momentum $L$. Their amplitudes allow, when compared to  Distorted Wave Born Approximation (DWBA) calculations, the determination of the neutron spectroscopic factor $C^2S$ values for the $(p,d)$ reaction, which are discussed first. 

DWBA calculations have been performed with the code FRESCO \cite{fresco} using the optical parameters given in \cite{Suppl2}. For \Capd,  three successive  angular distribution patterns are clearly identified as a function of increasing $E_x$: $L=2$ for the g.s. (Fig.~\ref{fig:XS}a)), $L=0$ for the three states at $E_x=3.06(2)$, 4.24(4) and 4.71~(9)~MeV (Figs.~\ref{fig:XS}b-d)), and $L=2$ for the four peaks at $E_x$ = 5.41, 6.54, 7.84 and 9.01~MeV (see Fig. 2 of \cite{Suppl3}). These distributions likely correspond to the removal of neutrons in $^{37}$Ca from the $1d_{3/2}$, $2s_{1/2}$ and $1d_{5/2}$ orbitals, leading to a sequence of expected $J^\pi=0^+$, ($1^+$ or $2^+$), and  ($1^+$ - $4^+$) states, respectively.

\begin{table*}[t]
\setlength\extrarowheight{1pt}
\caption{Summary of the experimental results and shell model calculations for the $^{36}$Ca - $^{36}$S mirror pair states with their proposed $J^\pi$ values, excitation energy $E_x$ in MeV, proton (neutron) spectroscopic factor values $C^2S$ from the $(p,d)$ reaction and mirror energy differences (MED) in keV. The $C^2S$ of the 0$^{+}_1$ state is obtained when  assuming a neutron removal from the  $d_{3/2}$ orbital, while for the 2$^{+}_1$, 1$^{+}$ and 2$^{+}_2$  states a removal from the  $s_{1/2}$ orbital is assumed. The last column shows the fraction of the wave function corresponding to the $0p-0h$ and $2p-2h$ configurations, according to the SM-CI calculations.} 

\begin{tabular}{ccccccccccccccccc}
\hline
\hline
\multirow{3}{*}{$J^\pi$}      & \multicolumn{5}{c}{$^{36}$Ca present work}                                                                                         &                      & \multicolumn{5}{c}{$^{36}$S}                                                                                          &                      & \multicolumn{2}{c}{MED}  \\
\cline{2-6} \cline{8-12} \cline{14-15} 
    & \multicolumn{2}{c}{Exp.}   && \multicolumn{2}{c}{Th.}    && \multicolumn{2}{c}{Exp. \cite{Gray70}}       && \multicolumn{2}{c}{Th.}   & & \multirow{2}{*}{Exp.} & \multirow{2}{*}{Th.} && $0p-0h$ / $2p-2h$ \\
\cline{2-3} \cline{5-6} \cline{8-9} \cline{11-12}
       & $E_x$   & $C^2S$   && $E_x$   & $C^2S$  && $E_x$ & $C^2S$  && $E_x$ & $C^2S$   &&  & &&\\ \hline
$0^+$  &    0.0     & 1.06(22) &&0.0         & 0.91    &&  0.0     & 1.06&&  0.0     &  0.93    &&  & &&0.92 / 0.08\\
$0^+_2$\footnotemark[1]\footnotemark[2]& 2.83(13)&          &&  2.70   &   $\simeq 0.0$      && 3.346 &         &&  3.42 &   $\simeq 0.0$       && -516(130) & -720 &&0.06 / 0.82\\
$2^+_1$& 3.06(2) & 0.66(14) &&  2.95   &1.01     && 3.291 & 0.86(17)&&  3.25 & 1.09     && -245(2)\footnotemark[3]& -300 &&0.79 / 0.20\\
$1^+_1$& 4.24(4) & 0.61(13) &&  4.00   &0.71     && 4.523 & 0.75(15)&&  4.22 & 0.72     && -280(41) & -220 
&&0.90 / 0.10\\
$2^+_2$& 4.71(9) & 0.28(7)  &&  3.81   &0.10     && 4.577 & 0.25(5) &&  4.54 & 0.05     && +133(90)       & -730 &&0.12 / 0.79\\
$(0^+_3)$\footnotemark[2]& 4.83(17)&          &&  4.36   &  $\simeq 0.0$       &&       &         &&  4.92 & $\simeq 0.0$    &&          & -560 &&0.87 / 0.12\\

\hline
 \hline
\end{tabular}
\label{tab:res}
\footnotetext[1] {Its configuration in $^{36}$Ca, according to the present SM-CI calculations, is protons $(d_{5/2})^{6-0.24}$  $(s_{1/2})^{2-0.48}$ $(d_{3/2})^{4-1.42}$ $(f_{7/2})^{0+1.8}$ $(p_{3/2})^{0+0.21}$ $(p_{1/2})^{0+0.02}$ $(f_{5/2})^{0+0.1}$ and neutrons $(d_{5/2})^{6-0.71}$  $(s_{1/2})^{2-0.85}$ $(d_{3/2})^{0+1.56}$. The first term in each superscript represents the normal occupancy value of the orbit, the second, preceded by a '+' / '-' sign  its calculated excess / reduction.}

\footnotetext[2]{State observed in the $^{38}$Ca($p$,$t$)$^{36}$Ca reaction. The obtained TNA, as well as that of the \textit{g.s.}, are given in \cite{Suppl4}.}
\footnotetext[3]{Value computed using the most precise energy of $E_x=3045(2)$~keV  for the 2$^+_1$  of $^{36}$Ca from Ref~\cite{Amth09}.}
\end{table*}

Spectroscopic factors, the uncertainties of which are dominated by the systematic error induced by the choice of optical potential parameters, are given in Table \ref{tab:res}. A $C^2S$ value  of 1.06(22) is found for the g.s.,  in excellent agreement with the value of 1.06 found in the mirror reaction \cite{Gray70} (see Table~\ref{tab:res}). Within the error bars,  it corresponds to the occupancy of the $1d_{3/2}$ orbital by about one neutron (proton) in $^{37}$Ca ($^{37}$Cl).  For the 2$^+_1$, 1$^+$ and 2$^+_2$ excited states, $C^2S$  values of 0.66(14), 0.61(13) and 0.28(7) have been found, respectively, also fully compatible with those obtained in the mirror reaction. With these three states, a large fraction of the $2s_{1/2}$ strength, $C^2S = 1.55(34)$,  has been collected.  At higher excitation energy, an integrated summed $C^2S$ value  of 4.38(88)  has been obtained from 5 to 9.5~MeV (see Fig.~2 of ~\cite{Suppl3}), to be compared to the full occupancy of the $1d_{5/2}$ orbital by 6 neutrons.

In the \Capt~ reaction, two states are observed in Fig.~\ref{fig:Ex}c), gated on Ca in the ZDD: the $0^+$ g.s. of $^{36}$Ca, as well as an excited state at 2.83(13)~MeV. Figs.~\ref{fig:XS}e,f) show the corresponding differential cross sections. The Two-Nucleon Amplitudes (TNA)  have been computed using Shell Model calculations with Configuration Interaction (SM-CI)   \cite{valiente}, for a transition from the g.s. of $^{38}$Ca to the 0$^+_1$, as well as to a 0$^+_2$ or 2$^+_1$ excited state in $^{36}$Ca. These TNA, given in \cite{Suppl4}, have been used to perform DWBA calculations which are  compared to experimental differential cross sections. As shown in Fig.~\ref{fig:XS}e), an excellent agreement is obtained for the ground state with an  $L=0$ shape, which is the only possibility for a $0^+\rightarrow0^+$ transition. The TNA that contributes by far the most to the reaction is the one arising from the removal of a pair of neutrons from the $1d_{3/2}$ orbital.

The shape of the angular distribution Fig.~\ref{fig:XS}f) of the 2.83(13)~MeV state,  is much better fitted as well when assuming an $L=0$  ($\chi^2$/ndf = 2.2/8 for a fit up to 40$^\circ$, green line) rather than an $L=2$ distribution ($\chi^2$/ndf = 13.1/8, blue line). It provides strong evidence of the existence of a 0$^+_2$ state below the 2$^+_1$ state, although a small contribution of the 2$^+_1$, lower than 20\%, could not be excluded. 

 Fig.~\ref{fig:Ex}d) has been obtained with a gate on outgoing Ar nuclei and $1\le E_p^{c.m.} (MeV)\le 2$ to select states in $^{36}$Ca that underwent a sequential $2p$ decay through the $1/2^+$ resonant state at 1.553(5)~MeV in $^{35}$K to the $^{34}$Ar g.s. Three peaks can be clearly identified at 4.83(17), 6.60(14) and 8.52(15)~MeV.  The decay of the relatively low excitation energy 4.83(17) MeV state through this low-$J$ resonance, will be strongly favored only if the state has $J^\pi$= $0^+$ or $1^+$, as the decay can proceed through the emission of an $L=0$ proton, contrary to the case of higher spin values. We tentatively propose a $J^\pi=0^+_3$ assignment to the 4.83(17) MeV state as the two-neutron transfer cross section for the odd-$J$, $1^+$ state, is predicted to be orders of magnitude lower. Moreover, a $J^\pi=0^+_3$ state at 4.967 MeV has been strongly populated in the $^{36}$Ar($p,t$)$^{34}$Ar reaction, involving isotone nuclei \cite{Pad72}.

\begin{figure}[h]
\begin{center}
\centering
\includegraphics[width=1.0\columnwidth]{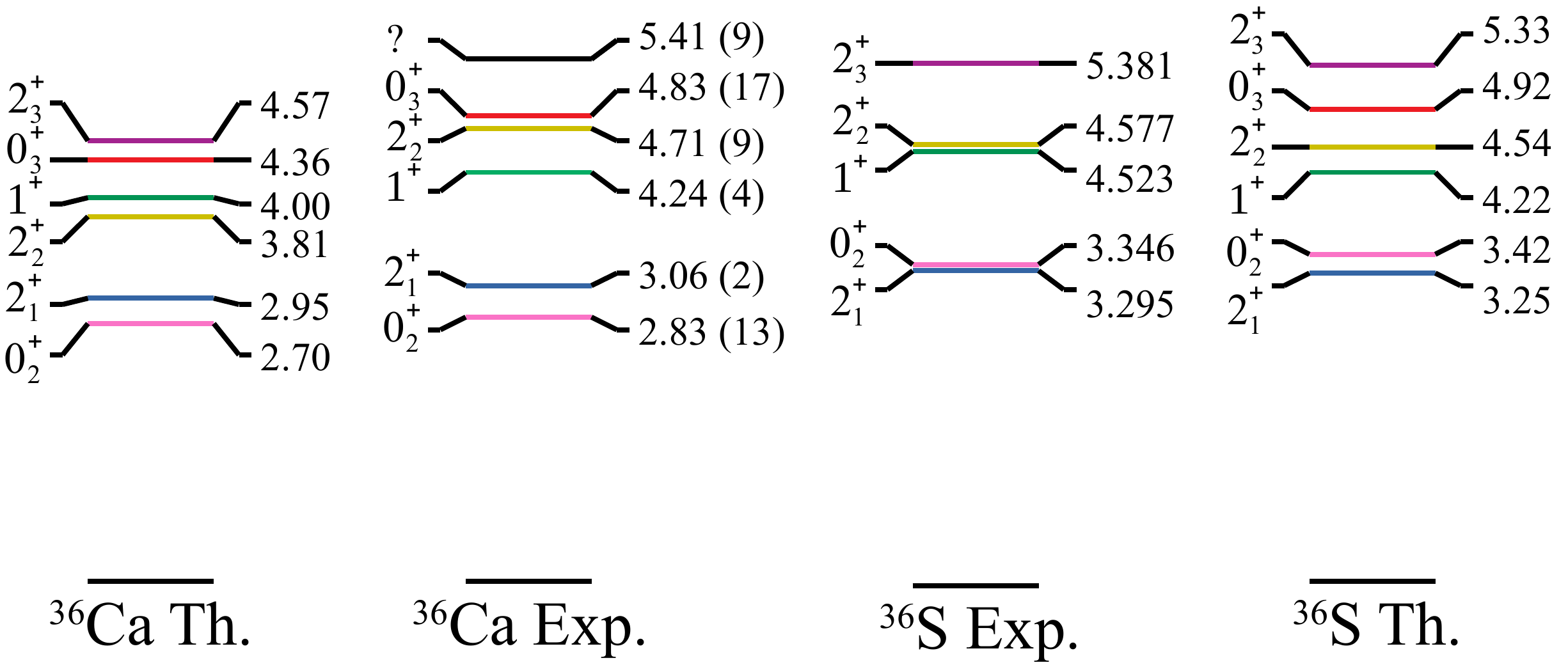}
\end{center}
\caption{Partial level scheme of the mirror pair $^{36}$Ca -$^{36}$S obtained experimentally (Exp.) and from Shell Model calculations (Th.). Values for $^{36}$Ca are those obtained in this work.}
\label{LevelScheme}
\end{figure}

\noindent {\sl Discussion.} The mirror pair $^{36}$Ca-$^{36}$S nuclei have been calculated with the shell-model code Antoine  \cite{rmp}  using the same valence space
     and interactions as in \cite{valiente}. There, the nuclear, isospin conserving part, is given by the $sdfpu$-mix interaction
     \cite{caurier2014}. The two-body matrix elements of the Coulomb interaction are computed with harmonic oscillator wave functions
     with \mbox{$\hbar \omega =  41 A^{-1/3} - 25 A^{-2/3}$}. The Coulomb corrections to the single-particle energies are taken from the
     experimental spectra of the $A$ = 17 and $A$ = 41 mirror nuclei.  Theoretical level schemes of the  mirror nuclei are compared to experimental ones in Fig.~\ref{LevelScheme}. The
     spectroscopic factors ($C^2S$) for the $(p,d)$ reaction  are gathered in Table~\ref{tab:res}.

Starting with $^{36}$S,  the correspondence between theory and experiment is very good (see  Fig.~\ref{LevelScheme}). The 0$^+_3 $ state, predicted at 4.92 MeV has not yet been observed experimentally. It is seen in the last column of Table~\ref{tab:res} that all states have well defined structures and, of the six lower states, two are calculated to be intruders. 

The present experiment confirms the energy of the 2$^+_1$ state in $^{36}$Ca \cite{Door07,Burg07}, which is, using the precise value of Ref.~\cite{Amth09},  245(2)~keV lower than in $^{36}$S. Populated by the ($p,d)$ reaction, it has a large $C^2S$ value with an $L=0$ angular distribution pattern, pointing to a pure $1p1h$ configuration with a hole in the $2s_{1/2}$ orbital. The  1$^+$ state has a similar structure and experiences a similar shift of -280(41) keV. These shifts can be attributed to their reduction of  $2s_{1/2}$ occupancy, as compared to the ground state. As protons in this orbital feel a smaller Coulomb repulsion than in the $d_{3/2}$ one, the 1$^+$, 2$^+$ excitation energies in $^{36}$S are pushed upward with respect to $^{36}$Ca. The calculated $C^2S$ value of the 2$^+_2$ in the two mirror nuclei, predicted to be an intruder state, is 3-5 times smaller than the 2$^+_2$ experimental value. This points to an  incorrect interpretation of its structure, which probably explains why experimental and calculated MEDs given in Table~\ref{tab:res} disagree. It is not excluded that the observed state would be in reality the 2$^+_3$ (with the 2$^+_2$ state unobserved). Further experimental and theoretical investigations are needed to elucidate this point.

The largest MED shift is for the 0$^+_2$, which is observed in $^{36}$Ca for the first time.  As discussed in \cite{valiente}, its very large MED of -516(130) keV, comparable to the calculated value of -720 keV,  is due to the addition of two contributions. The first comes from its $2p2h$ proton intruder nature, because the two protons promoted across $Z=20$ in the $pf$ shell feel less Coulomb repulsion than in the $sd$ shell. Moreover, the opening of the proton core leads to an increase of the degree of collectivity and has a large influence on the neutron configuration, understood to be a $1p1h$ with one neutron missing in the  $2s_{1/2}$ orbital, as for the (2$^+_1$, 1$^+$) doublet. These two effects stem naturally from the occupancies (listed below Table~\ref{tab:res} in the item $^{a)}$) and sum coherently to generate the very large observed MED.

Concerning the  $0^+_3$ state, its structure is predicted to be dominated by two neutron holes in the $2s_{1/2}$ orbital, which would result in a large central depletion similar to that described for protons in  \cite{Muts17}. A Coulomb shift of -560 keV, twice as large as for the $2^+_1$ and $1^+_1$ states, is expected from the calculation. In $^{36}$S an unobserved $0^+_3$ state, should be present around 4.9~MeV. 

\noindent {\sl Summary-} The $^{37}$Ca$(p,d)^{36}$Ca transfer reaction was used to populate the  ground state of $^{36}$Ca as well as the ($2^+_{1,2}$, $1^+_1$) states, through the removal of a neutron from the $1d_{3/2}$ and $2s_{1/2}$ orbitals, respectively. Their $L$ assignments and $C^2S$ values are deduced from the comparison between their experimental differential cross section and DWBA calculations. Very similar $C^2S$ values were found with the quasi-mirror reaction $^{37}$Cl$(d,^3$He$)^{36}$S for the states up to an excitation energy of 5 MeV, pointing to a very similar structure between the mirror states of $^{36}$Ca  and $^{36}$S. The large observed MED, of about $-250$~keV for the $2^+_{1}$ and $1^+_1$ states, is understood as originating from their pure $1p1h$ structure, with one neutron (proton) less in the  $2s_{1/2}$ orbital, as compared to the ground state. The  $^{38}$Ca$(p,t)^{36}$Ca reaction revealed the existence of a 0$^+_2$ state at 2.83(13) MeV, that would correspond to the ground state intruder configuration in $^{32}$Ca, if mirror symmetry with $^{32}$Mg would be preserved. This 0$^+_2$ experiences a spectacular MED of about -516(130) keV, which is interpreted by its combined neutron $1p1h$ and proton $2p2h$ intruder components.  This amazingly large isospin-symmetry breaking is extremely rare in the chart of nuclides. It  can be fully interpreted by the effect of the Coulomb interaction and is favored because of the shape coexistence between the two 0$^+$ states in $^{36}$Ca. It makes the $^{36}$Ca-$^{36}$S mirror pair a remarkable physics case in which the Coulomb interaction acts as a magnifying glass to probe their structures, without perturbing them.


The present work, together with the ground state properties of $^{36}$Ca \cite{Surb21,Miller19}, is expected to serve as a benchmark case for ab-initio calculations that are supposed to rigorously treat all ISB effects of the nuclear force. In the broader context of nuclear astrophysics (and in particular for the rp-process), the present conclusions showing good symmetry in $C^2S$ strengthen the validity of using the same $C^2S$ values between mirror reactions, even when large MED are present, to determine unknown reaction cross sections.

\acknowledgments {\small
The continued support of the staff of the GANIL facility is gratefully acknowledged. DS was supported by the JSPS KAKENHI Grant Number 19H01914. AP's work is supported in part by the Ministerio de Ciencia, Innovaci\'on y Universidades (Spain), Grant CEX2020-001007-S  funded by MCIN/AEI/10.13039/501100011033
 and grant PGC-2018-94583. Support from the NFS grant PHY-1811855 is also acknowledged.}

\end{document}


\title{Supplemental Material for "The structure of $^{36}$Ca under the Coulomb magnifying glass"}

\author{L.~Lalanne}
    \email[]{louis.lalanne@ijclab.in2p3.fr}
    \affiliation{Universit\'e Paris-Saclay, CNRS/IN2P3, IJCLab, 91405 Orsay, France}
    \affiliation{Grand Accélérateur National d'Ions Lourds (GANIL), CEA/DRF-CNRS/IN2P3, 
    Bd. Henri Becquerel, 14076 Caen, France}  
\author{O.~Sorlin}
    \email[]{olivier.sorlin@ganil.fr}
    \affiliation{Grand Accélérateur National d'Ions Lourds (GANIL), CEA/DRF-CNRS/IN2P3, 
    Bd. Henri Becquerel, 14076 Caen, France}
    \author{A.~Poves}
    \affiliation{Departamento de F\'isica Te\'orica and IFT-UAM/CSIC, Universidad Aut\'onoma de Madrid,  E-2804 Madrid, Spain}
\author{M.~Assi\'e}
    \affiliation{Universit\'e Paris-Saclay, CNRS/IN2P3, IJCLab, 91405 Orsay, France}
\author{F.~Hammache}
    \affiliation{Universit\'e Paris-Saclay, CNRS/IN2P3, IJCLab, 91405 Orsay, France}
\author{S.~Koyama}
    \affiliation{Department of Physics, Unviversity of Tokyo}
     \affiliation{Grand Accélérateur National d'Ions Lourds (GANIL), CEA/DRF-CNRS/IN2P3, 
    Bd. Henri Becquerel, 14076 Caen, France}
\author{F.~Flavigny}
    \affiliation{Normandie Univ, ENSICAEN, UNICAEN, CNRS/IN2P3, LPC Caen, 14000 Caen, France}
\author{V.~Girard-Alcindor}
    \affiliation{Grand Accélérateur National d'Ions Lourds (GANIL), CEA/DRF-CNRS/IN2P3, 
    Bd. Henri Becquerel, 14076 Caen, France}
    \author{A.~Lemasson}
    \affiliation{Grand Accélérateur National d'Ions Lourds (GANIL), CEA/DRF-CNRS/IN2P3, 
    Bd. Henri Becquerel, 14076 Caen, France}
  \author{A.~Matta}
       \affiliation{Normandie Univ, ENSICAEN, UNICAEN, CNRS/IN2P3, LPC Caen, 14000 Caen, France}
       \author{T.~Roger}
    \affiliation{Grand Accélérateur National d'Ions Lourds (GANIL), CEA/DRF-CNRS/IN2P3, 
    Bd. Henri Becquerel, 14076 Caen, France}
\author{D.~Beaumel}
    \affiliation{Universit\'e Paris-Saclay, CNRS/IN2P3, IJCLab, 91405 Orsay, France}
\author{Y~Blumenfeld}
    \affiliation{Universit\'e Paris-Saclay, CNRS/IN2P3, IJCLab, 91405 Orsay, France}
\author{B.~A.~Brown}
 \affiliation{Department of Physics and Astronomy, National Superconducting Cyclotron Laboratory,
Michigan State University, East Lansing, Michigan}
\author{F.~De Oliveira Santos}
 \affiliation{Grand Accélérateur National d'Ions Lourds (GANIL), CEA/DRF-CNRS/IN2P3,
    Bd. Henri Becquerel, 14076 Caen, France}
\author{F.~Delaunay}
    \affiliation{Normandie Univ, ENSICAEN, UNICAEN, CNRS/IN2P3, LPC Caen, 14000 Caen, France}
    \author{N.~de~S\'{e}r\'{e}ville}
    \affiliation{Universit\'e Paris-Saclay, CNRS/IN2P3, IJCLab, 91405 Orsay, France}
\author{S.~Franchoo}
    \affiliation{Universit\'e Paris-Saclay, CNRS/IN2P3, IJCLab, 91405 Orsay, France}
\author{J.~Gibelin}
    \affiliation{Normandie Univ, ENSICAEN, UNICAEN, CNRS/IN2P3, LPC Caen, 14000 Caen, France}
\author{J.~Guillot}
    \affiliation{Universit\'e Paris-Saclay, CNRS/IN2P3, IJCLab, 91405 Orsay, France}
\author{O.~Kamalou}
    \affiliation{Grand Accélérateur National d'Ions Lourds (GANIL), CEA/DRF-CNRS/IN2P3, 
    Bd. Henri Becquerel, 14076 Caen, France}
\author{N.~Kitamura}
    \affiliation{Center for Nuclear Study, University of Tokyo}

\author{V.~Lapoux}
    \affiliation{CEA, Centre de Saclay, IRFU, Service de Physique Nucléaire, 91191 Gif-sur-Yvette, France}

\author{B.~Mauss}
    \affiliation{RIKEN Nishina Center, 2-1, Hirosawa, Wako, Saitama 351-0198, Japan}
    \affiliation{Grand Accélérateur National d'Ions Lourds (GANIL), CEA/DRF-CNRS/IN2P3, 
    Bd. Henri Becquerel, 14076 Caen, France}
\author{P.~Morfouace}
    \affiliation{Grand Accélérateur National d'Ions Lourds (GANIL), CEA/DRF-CNRS/IN2P3, 
    Bd. Henri Becquerel, 14076 Caen, France}
    \affiliation{CEA, DAM, DIF, F-91297 Arpajon, France}
\author{M.~Niikura}
    \affiliation{Department of Physics, Unviversity of Tokyo}
\author{J.~Pancin}
    \affiliation{Grand Accélérateur National d'Ions Lourds (GANIL), CEA/DRF-CNRS/IN2P3, 
    Bd. Henri Becquerel, 14076 Caen, France}

\author{T.~Y.~Saito}
    \affiliation{Department of Physics, Unviversity of Tokyo}
\author{D.~Suzuki}
    \affiliation{RIKEN Nishina Center, 2-1, Hirosawa, Wako, Saitama 351-0198, Japan}
\author{C.~Stodel}
    \affiliation{Grand Accélérateur National d'Ions Lourds (GANIL), CEA/DRF-CNRS/IN2P3, 
    Bd. Henri Becquerel, 14076 Caen, France}
\author{J-C.~Thomas}
    \affiliation{Grand Accélérateur National d'Ions Lourds (GANIL), CEA/DRF-CNRS/IN2P3, 
    Bd. Henri Becquerel, 14076 Caen, France}

\date{\today}
\maketitle

\section{S.1 Phase Space}

\begin{figure}[h]
\begin{center}
\centering

\includegraphics[height=0.47\columnwidth]{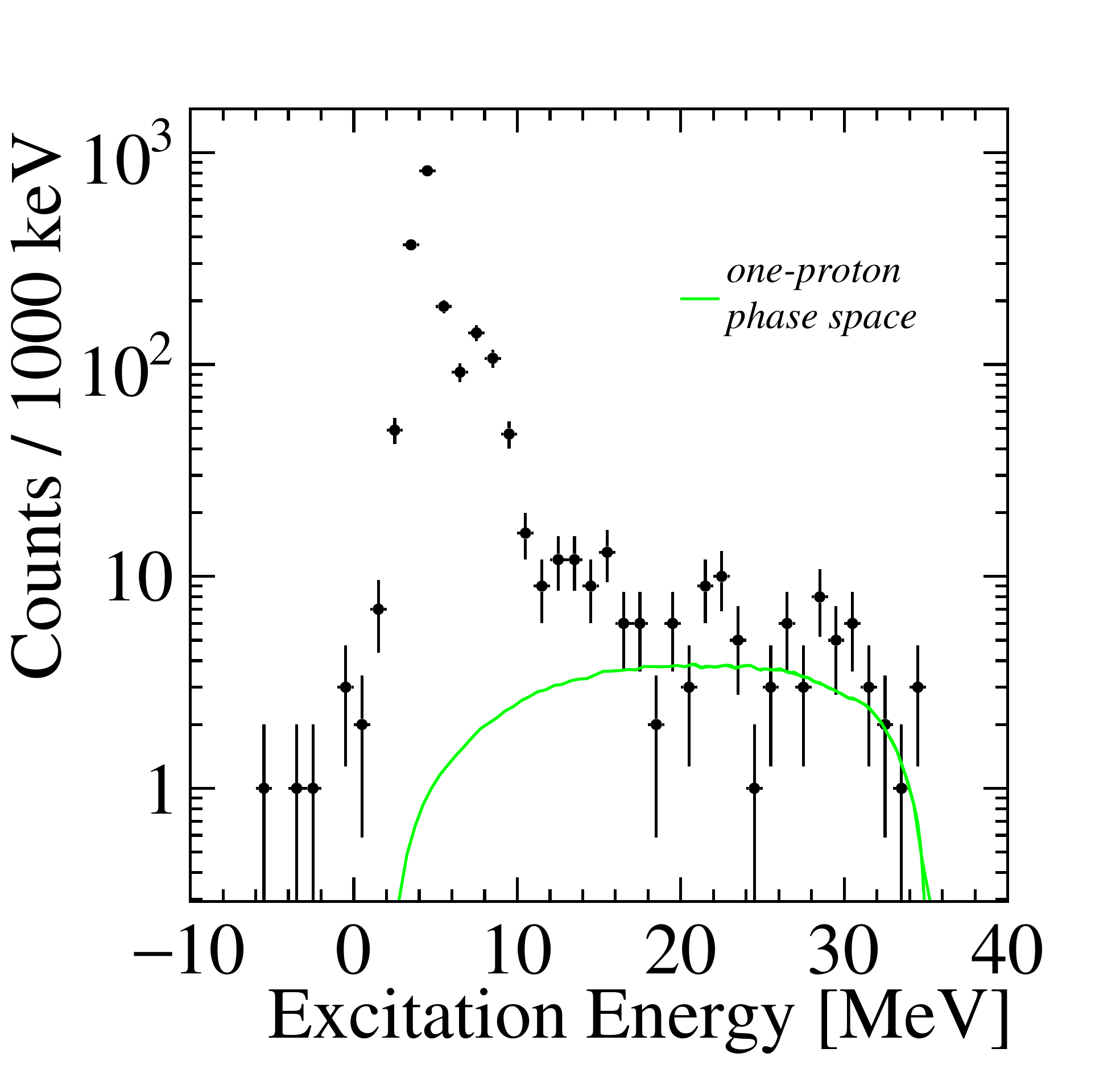}
\includegraphics[height=0.45\columnwidth]{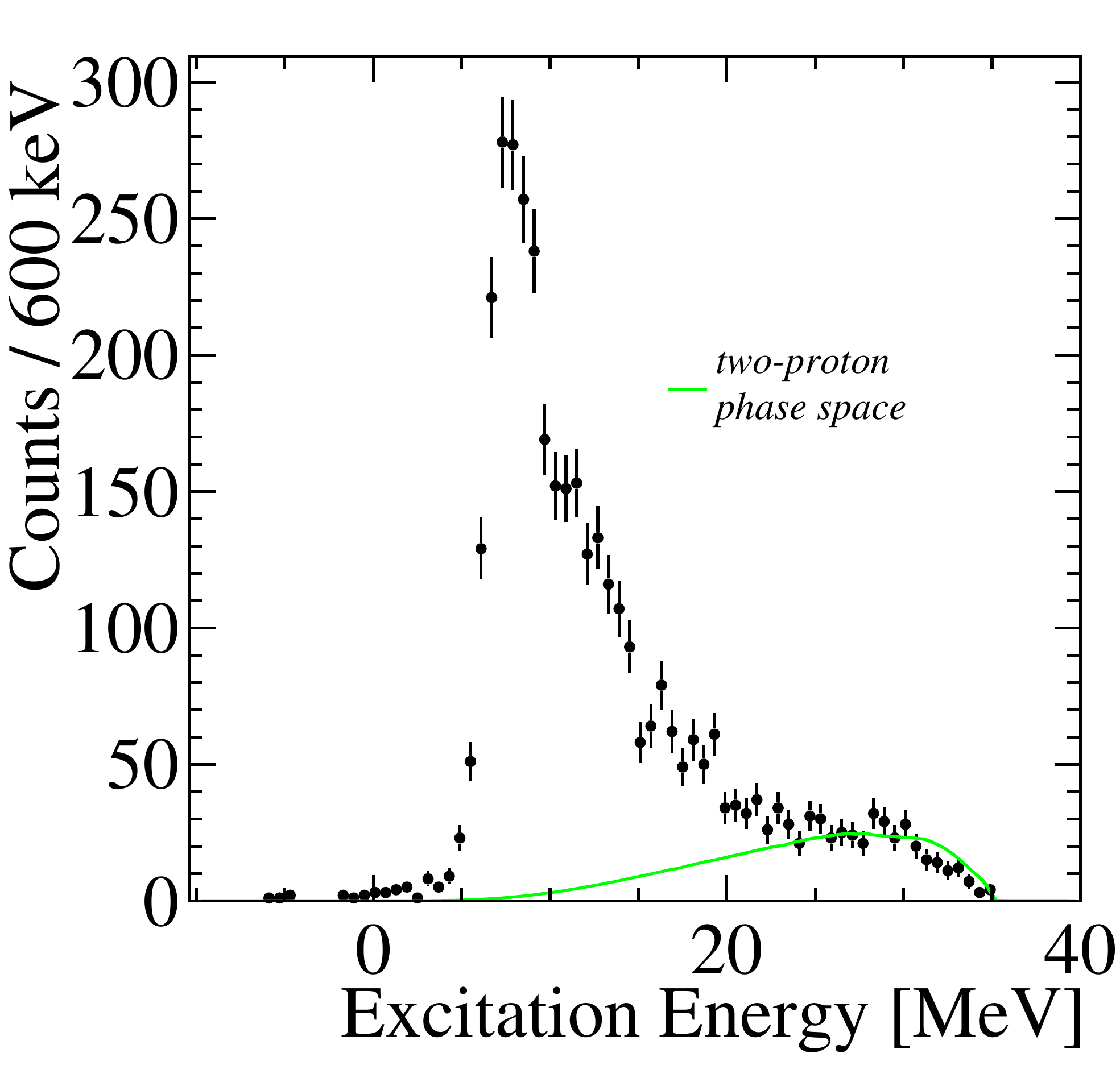}
\end{center}
\caption{Excitation energy spectrum of $^{36}$Ca reconstructed using the $^{37}$Ca($p,d)^{36}$Ca reaction and obtained with a gate on outgoing K (left) and Ar (right) isotopes. The green line shows the computed one (left) and two-proton phase space (right), normalized in order to reproduce the data between 25 and 40~MeV}
\label{fig:PS}
\end{figure}

The contribution of the phase space to the excitation energy spectrum has been investigated. The shapes of the one and two-proton phase spaces have been calculated using the TGenPhaseSpace class of ROOT and corrected from geometrical efficiency determined using NPTool simulations \cite{Matta2016}. Fig.~\ref{fig:PS} shows the excitation energy spectrum of $^{36}$Ca reconstructed using the $^{37}$Ca($p,d)^{36}$Ca reaction. A gate is applied on outgoing K (left of Fig.~\ref{fig:PS}) or Ar isotopes (right of Fig.~\ref{fig:PS}) in order to isolate the one or two-proton decay channel, respectively. The green line shows the one or two-proton phase space, normalized in order to reproduce the data between 25 and 40~MeV. Since the intensity of the phase space is maximum at about 25-30~MeV, this allows to determine the maximum contribution of the phase spaces in the energy range of interest (between 0 and 10 MeV). Its maximum contribution is found to be about 1\% for the one-proton phase space and about 2\% for the two-proton one at 10~MeV and decreases when going to lower excitation energy, allowing to neglect phase-space contributions in our work.  

\section{S.2 DWBA analysis}

Calculation of the differential cross sections have been performed within the Distorded Wave Born Approximation (DWBA). The prescriptions used for the DWBA analysis of the \Capd~reaction are:
\begin{itemize}
\item Entrance potential describing the p + $^{37}$Ca channel: Menet $et~al.$ global proton optical potential \cite{Menet71}.
\item Exit potential describing the d + $^{36}$Ca channel: finite range adiabatic potential following the Jonhson-Tandy prescription \cite{Jonhson74} and using the Becchetti and Greenlees global optical potentials \cite{BG69} for the proton and the neutron inside the deuteron.
\item The bound state wave function of the neutron in $^{37}$Ca has been computed using a Wood-Saxon potential using a radius r$_0 = 1.27$~fm, a diffusivity a$_0=0.67$~fm and a spin-orbit potential V$_{SO}=6$~MeV (following the prescriptions of Ref.~\cite{BM98}). The depth of the potential well has been adjusted to reproduce the binding energy of the orbital from which the neutron was removed.
\item The bound state wave function of the neutron in the deuteron has been computed using a classical Reid soft-core potential \cite{Reid}. The $\bra{p}\ket{d}$ vertex has been treated in finite range.
\end{itemize} 

In the case of the $^{38}$Ca($p$,$t$)$^{36}$Ca reaction, the same prescriptions have been used except for the exit channel potential for which the global triton potential of Ref.~\cite{Li07} has been used. Two-neutron transfer calculations include both direct and sequential transfer. For the latter, no experimental information was used for the intermediate nucleus $^{37}$Ca (the binding energy of intermediate nucleus is taken as the mean of the binding
energies of initial and final nuclei and no intermediate resonance is considered).

\section{S.3 The $d_{5/2}$ states}

\begin{figure}[]
\begin{center}
\centering
\includegraphics[width=0.95\columnwidth]{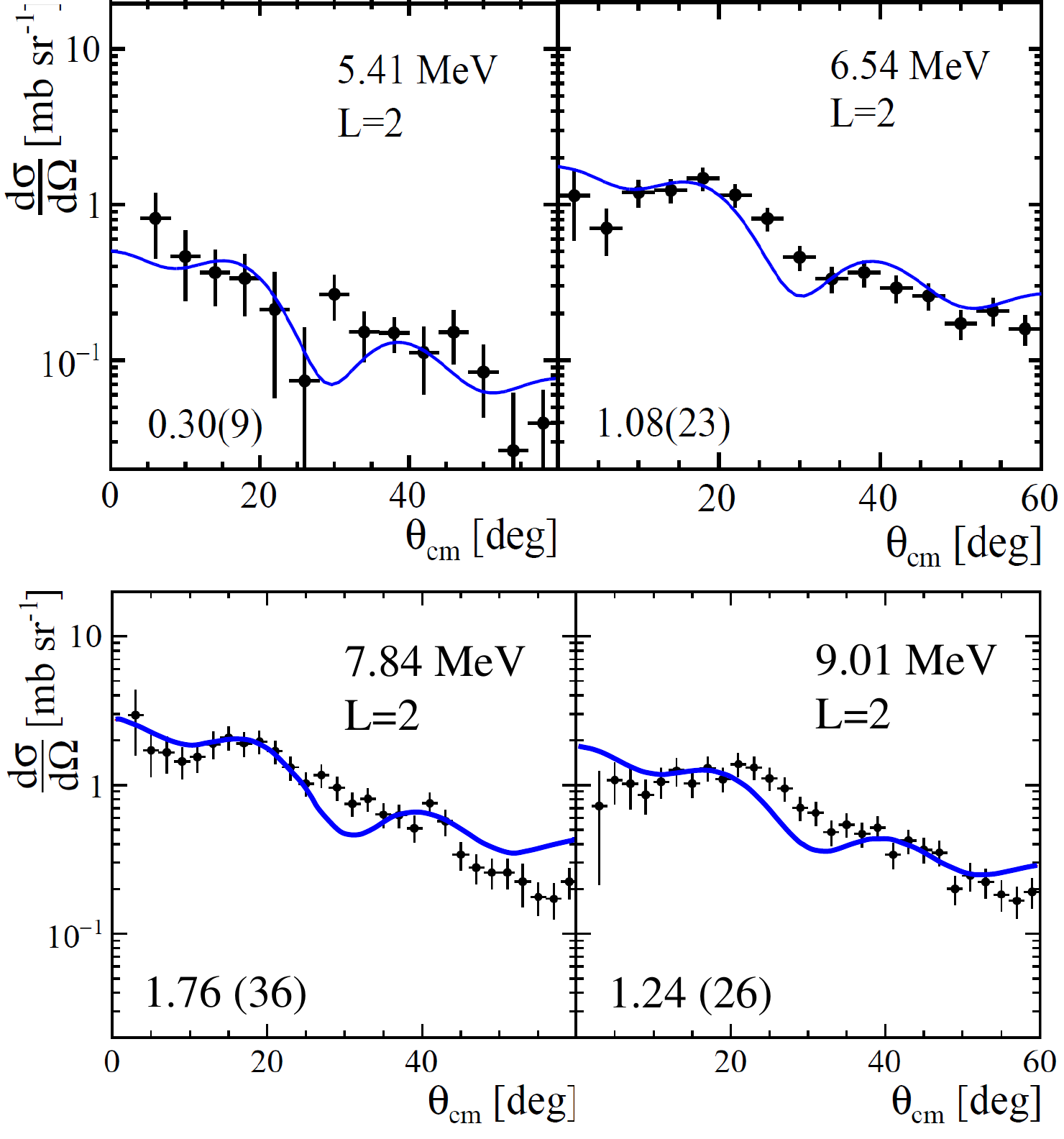}

\end{center}
\caption{Differential cross sections obtained for the states at 5.41, 6.54, 7.84 and 9.01 MeV. The blue lines show the results of the DWBA calculations, using an $L=2$ neutron transfer. Corresponding spectroscopic factors are given in the left bottom corner of each plot.}
\label{fig:XS8MeV}
\end{figure}

Between 5 and 10~MeV, a large cross section is observed in the excitation energy spectrum of $^{36}$Ca, using the $^{37}$Ca($p$,$d$)$^{36}$Ca reaction (see Fig.~1 of the letter). A high density of state is expected to be populated in this energy range, with spin-parity from 1$^+$ to 4$^+$, from the removal of a $1d_{5/2}$ neutron in $^{37}$Ca. The fit of the excitation energy spectrum does not allow to conclude about the exact energy of the states due to the limited experimental resolution. However, the differential cross sections of Fig.~\ref{fig:XS8MeV} have been obtained by fixing the centroid of the Gaussian function at $E_x$ = 5.41, 6.54, 7.84 and 9.01~MeV in the fit, where peaks seem to be present. They all show a clear $L=2$ pattern, confirming the hypothesis of a $1d_{5/2}$ neutron hole configuration of these two resonances. With this four contributions, a large fraction of the $d_{5/2}$ strength has been collected, with an integrated $C^2S$ value of 4.38(88).   

\begin{figure}[]
\centering
        \includegraphics[width=\columnwidth]{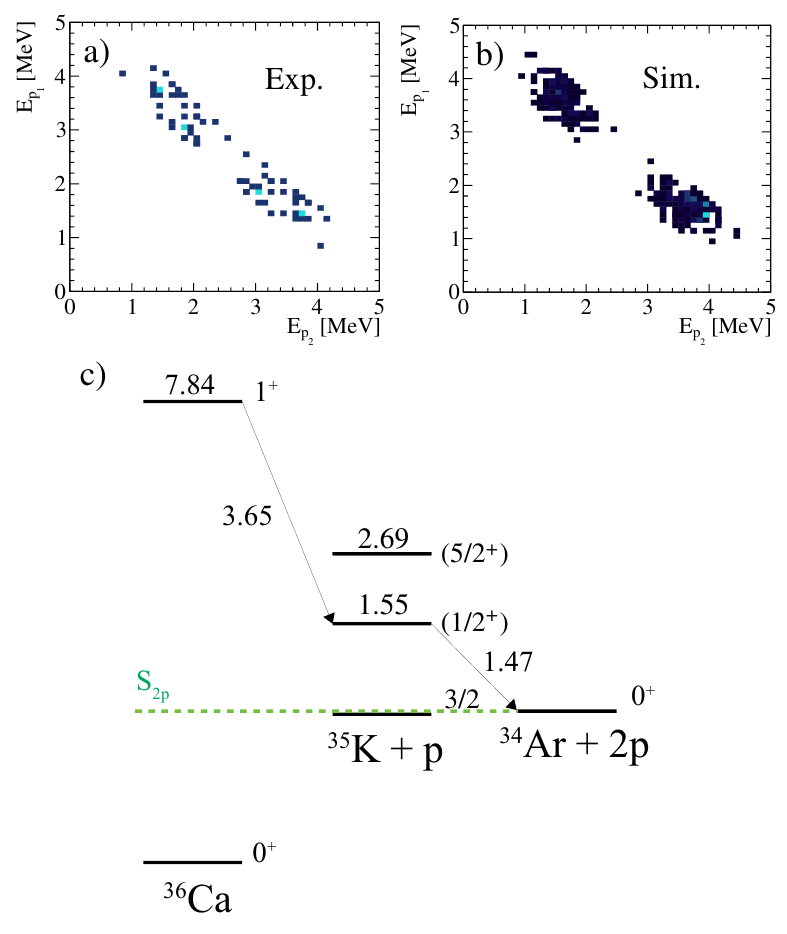}

\caption{Observed a) and simulated  b) correlations between the energy of the first and second proton emitted during the two-proton decay of the 7.8~MeV state through the 1/2$^+$ resonant state in $^{35}$K, as shown in the decay scheme of c).}
\label{fig:2p}
\end{figure} 

The two-proton decay pattern of the 7.8 MeV resonance can be studied to get insight in its spin-parity value. Fig.~\ref{fig:2p} shows the correlation between the energy of the first and second emitted protons during its decay. It has been obtained by selecting an outgoing Ar in the ZDD, as well as a deuteron and 2 protons in MUST2, at an excitation energy between 7.5 and 8.1~MeV. Two clear spots  are identified in Fig.~\ref{fig:2p}, at proton energies of 1.7(3)~MeV and 3.4(3)~MeV, in good agreement with the expected energies of 1.47 and 3.65~MeV arising from the sequential decay through the $1/2^+$ state at 1.55~MeV in $^{35}$K (see bottom part of Fig.~\ref{fig:2p}). This two-proton decay pattern is further confirmed by the simulation shown in the top right part of Fig.~\ref{fig:2p}.

As shown in Fig.~\ref{fig:2p}, the two-proton decay can either occur through the 1.55~MeV or 2.69~MeV state (higher ones are not considered here since they will not be favored). The 1/2$^+$ and 5/2$^+$ are based on the systematics along the K isotopic chain, and energies observed in the mirror nucleus $^{35}$S. One can compute the ratio of penetration barrier factors to the 1/2$^+$ and to the 5/2$^+$ states, for different spin-parity assumptions of the initial resonance. Among the possible  spin-parity values expected to be populated, the  observed decay pattern is by far most compatible with an $J^\pi=1^+$ initial state in $^{36}$Ca at 7.8~MeV, which decays with an $L=0$ proton decay to the 1/2$^+$ resonance in $^{35}$K in competition with an $L=2$, one-proton decay emission, to the ground state of $^{35}$K.

\section{S.4 Two Nucleon Amplitude}

The Two-Nucleon Amplitudes (TNA) for a transition from the ground state of $^{38}$Ca to the 0$^+_1$ ground state, as well as the 0$^+_2$ and 2$^+_1$ excited states of $^{36}$Ca, have been computed using the 
 Shell Model with Configuration Interaction (SM-CI) calculations. These TNA, given in Table~\ref{tab:TNA}, are used to perform the DWBA calculations of the theoretical differential cross section for the $^{38}$Ca($p$,$t$)$^{36}$Ca transfer reaction. 

\begin{table}[h!]
\centering
\caption{Two-nucleon amplitudes for the ground state, 0$^+_2$ state and 2$^+$ state of $^{36}$Ca from the two SM calculations.}

\label{tab:TNA}
\begin{tabular}{cccc}
\hline
\hline
                     & $(nlj)_1$ & $(nlj)_2$ & TNA                \\ \hline
\multirow{3}{*}{g.s.}  & $d_{3/2}$      & $d_{3/2}$             &    -0.97             \\
                     & $s_{1/2}$        & $s_{1/2}$             &    -0.21                \\
                     & $d_{5/2}$        & $d_{5/2}$             &    -0.20                 \\ \hline
\multirow{3}{*}{0$^+_2$} & $d_{3/2}$    & $d_{3/2}$             &    -0.05                   \\
                     & $s_{1/2}$        & $s_{1/2}$             &     0.29                  \\
                     & $d_{5/2}$        & $d_{5/2}$             &    -0.12                 \\ \hline
\multirow{5}{*}{2$^+$}  & $d_{3/2}$     & $d_{3/2}$             &     0.09                 \\
                     & $d_{3/2}$        & $s_{1/2}$             &    -1.42                 \\
                     & $d_{3/2}$        & $d_{5/2}$             &    -0.05                  \\
                     & $s_{1/2}$        & $d_{5/2}$             &    0.36                  \\
                     & $d_{5/2}$        & $d_{5/2}$             &     0.12                 \\ \hline
                     \hline
\end{tabular}
\end{table}

The main contributing channel for the ground state is the removal of a pair of neutrons from the $d_{3/2}$ orbital, in agreement with the expected structure of this state, while for the second 0$^+_2$ state, the removal of neutrons from the $s_{1/2}$ orbital dominates the total cross section. Both channels lead to  $L=0$ distributions. In the case of the 2$^+$ state, the cross section takes its major contribution from the removal of a neutron from the $d_{3/2}$ and another from the $s_{1/2}$ orbital. In this case an L=2 angular distribution would be observed.

\begin{figure}[h]
\begin{center}
\centering
\includegraphics[width=0.9\columnwidth]{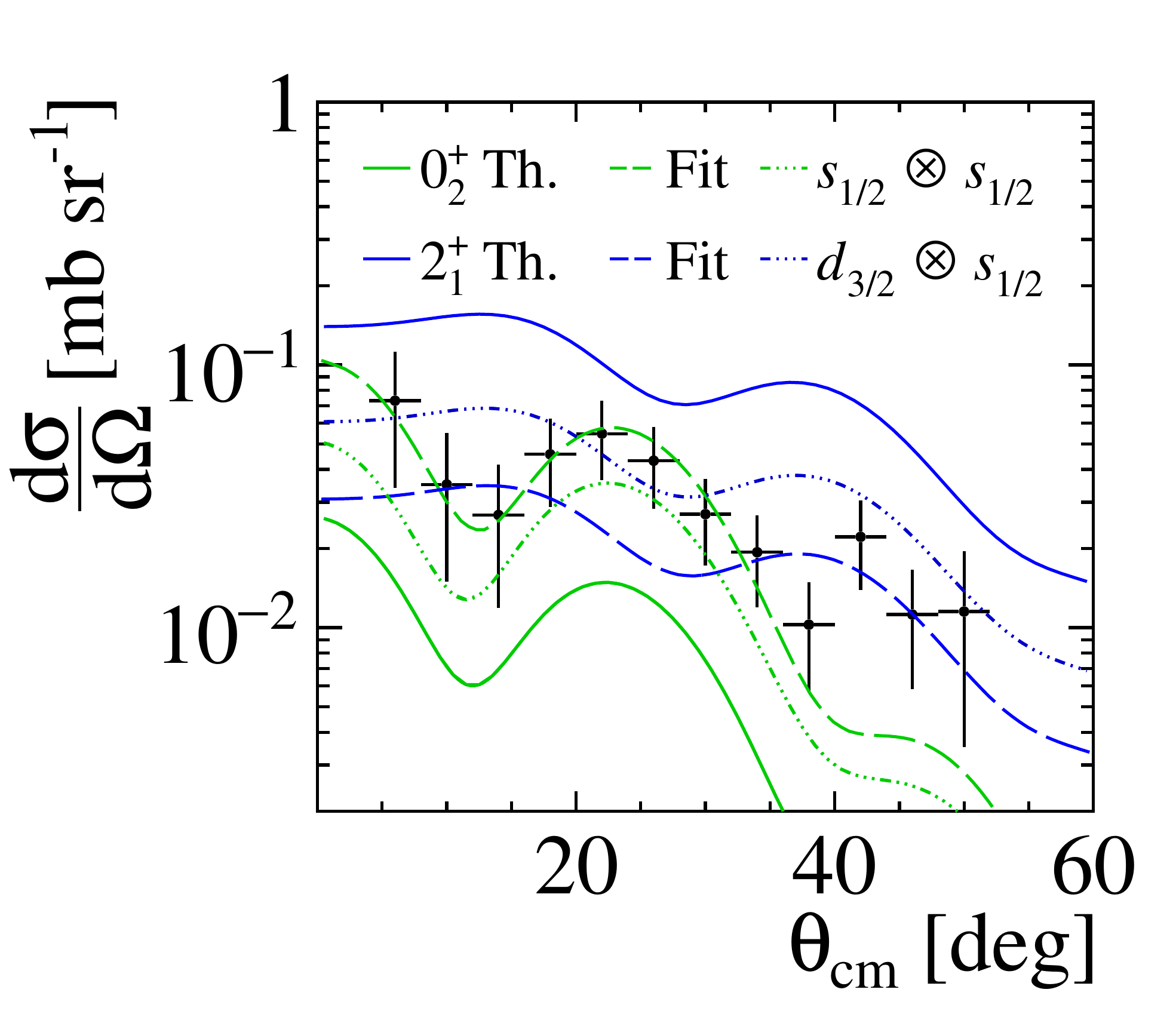}
\end{center}
\caption{Differential cross section obtained for the 0$^+_2$ intruder state. Black points are experimental data while the blue and green lines show the results of the DWBA calculations (using TNA values given in Table~\ref{tab:TNA}) for the 0$^+_2$ state and the 2$^+$ state, respectively. The full lines correspond to the result obtained considering all the contributing channels, dashed lines to the best fit to data, while the dashed-dotted line  shows the calculated values using the dominant TNA contribution only (see text for details).
}
\label{fig:XSTh}
\end{figure} 

The amplitude of the cross section of the 0$^+_2$ state is underestimated by the calculation, as shown by the full green line in Fig.~\ref{fig:XSTh}. However, by taking only into account the dominating channel in the calculation (removal of a pair of neutron in the $s_{1/2}$ orbital, see Table~\ref{tab:TNA} ), the amplitude is better reproduced (see the dashed-dotted green line in  Fig.~\ref{fig:XSTh}), highlighting the fact that the small TNA values of the competing channels generate destructive interferences with the dominant channel and affect significantly the amplitude of the cross section. At the opposite, the 2$^+$ cross section amplitude is largely over estimated by the calculation (blue dashed line) since no significant  $L=2$ component is needed to fit the data. Considering the dominant channel with one neutron in the $s_{1/2}$ and one in the $d_{3/2}$ orbital only, the theoretical cross section is significantly reduced, as shown with the dashed-dotted blue line in  Fig.~\ref{fig:XSTh}.
